\documentclass{article}

     \usepackage[preprint]{neurips_2019}

\usepackage[utf8]{inputenc} % allow utf-8 input
\usepackage[T1]{fontenc}    % use 8-bit T1 fonts
\usepackage{hyperref}       % hyperlinks
\usepackage{url}            % simple URL typesetting
\usepackage{booktabs}       % professional-quality tables
\usepackage{amsfonts}       % blackboard math symbols
\usepackage{nicefrac}       % compact symbols for 1/2, etc.
\usepackage{microtype}      % microtypography
\usepackage{graphicx}
\usepackage{xspace}
\usepackage{amsmath}
\usepackage{amssymb}
\usepackage{multirow}
\usepackage{adjustbox}
\usepackage{listings}
\usepackage{subcaption} %% For complex figures with subfigures/subcaptions
\usepackage{tikz}
\usepackage{flushend}
\usepackage{textcomp}
\usepackage{floatrow}
\newfloatcommand{capbtabbox}{table}[][\FBwidth]

\setcitestyle{numbers,,open={[},close={]}}

\newcommand{\etal}{\hbox{\emph{et al.}}\xspace}
\newcommand{\eg}{\hbox{\emph{e.g.}}\xspace}
\newcommand{\ie}{\hbox{\emph{i.e.}}\xspace}

\newcommand{\wrt}{\hbox{\emph{w.r.t.}}\xspace}
\newcommand{\etc}{\hbox{\emph{etc.}}\xspace}
\newcommand{\resp}{\hbox{\emph{resp.}}\xspace}

\newcommand{\dypro}{\textsc{DyPro}\xspace}

\title{Learning Scalable and Precise Representation \\ of Program Semantics}

\author{%
	Ke Wang \\%\thanks{Use footnote for providing further information about author (webpage, alternative address)---\emph{not} for acknowledging funding agencies.} \\
	Visa Research \\
	Palo Alto, CA 94306 \\
	\texttt{kewang@visa.com} \\
}

\makeatletter
\newenvironment{btHighlight}[1][]
{\begingroup\tikzset{bt@Highlight@par/.style={#1}}\begin{lrbox}{\@tempboxa}}
	{\end{lrbox}\bt@HL@box[bt@Highlight@par]{\@tempboxa}\endgroup}

\newcommand\btHL[1][]{%
	\begin{btHighlight}[#1]\bgroup\aftergroup\bt@HL@endenv%
	}
	\def\bt@HL@endenv{%
	\end{btHighlight}%   
	\egroup
}
\newcommand{\bt@HL@box}[2][]{%
	\tikz[#1]{%
		\pgfpathrectangle{\pgfpoint{1pt}{0pt}}{\pgfpoint{\wd #2}{\ht #2}}%
		\pgfusepath{use as bounding box}%
		\node[anchor=base west, fill=orange!25,outer sep=.5pt,inner xsep=0.5pt, inner ysep=-0.3pt, rounded corners=2pt, minimum height=\ht\strutbox-.1pt,#1]{\raisebox{.01pt}{\strut}\strut\usebox{#2}};
	}%
}
\makeatother

\definecolor{codegreen}{rgb}{0,0.6,0}
\definecolor{codegray}{rgb}{0.5,0.5,0.5}
\definecolor{codepurple}{rgb}{0.58,0,0.82}

\lstdefinestyle{mystyle}{
	commentstyle=\color{codegreen},
	keywordstyle=\color{blue}\bfseries,
	numberstyle=\tiny\color{codegray},
	stringstyle=\color{codepurple},
	basicstyle=\linespread{0.85}\fontsize{5.5}{8.8}\ttfamily,
	breakatwhitespace=false,         
	breaklines=false,                 
	captionpos=b,                    
	keepspaces=true,                 
	numbers=none,                    
	numbersep=3pt,                  
	showspaces=false,                
	showstringspaces=false,
	showtabs=false,                  
	tabsize=2,
	language=[Sharp]C,
	moredelim=**[is][\btHL]{`}{`},
	moredelim=**[is][{\btHL[fill=red!40]}]{@}{@},%,draw=red,dashed,thick	
}

\begin{document}

\maketitle

\begin{abstract}
	Neural program embedding has shown potential 
	in aiding the analysis of large-scale, complicated software. 
	Newly proposed deep neural architectures pride themselves 
	on learning program semantics rather than superficial 
	syntactic features. However, by considering the source 
	code only, the vast majority of neural networks do not 
	capture a deep, precise representation of program 
%	uncovering the deeper semantic properties 
%	that may require rigorous formal approaches 
	semantics. In this paper, we present \dypro, a novel deep neural network 
	that learns from program execution traces. Compared to 
	the prior dynamic models, not only is \dypro capable of 
	generalizing across multiple executions for learning 
	a program's dynamic semantics in its entirety, but \dypro 
	is also more efficient when dealing with programs yielding long execution 
	traces. For evaluation, we task \dypro with semantics classification 
	(\ie categorizing programs based on their semantics) 
	and compared it against two prominent static models: Gated Graph Neural 
	Network and TreeLSTM.  
	We find that \dypro achieves the highest prediction accuracy among 
	all models. To further reveal the capacity of all aforementioned deep neural 
	architectures, we examine if the models can learn to detect deeper semantic 
	properties of a program. In particular given a task of recognizing loop 
	invariants, we find that \dypro outperforms all static models by a wide margin.

%	
%	To further reveal the model capacity, a more challenging 
%	assignment --- semantic analysis --- is created for models to tackle. 
%	In view of the results, we conduct an 
%	additional study to further shed light on the limitations of 
%	static models. 	
%	In particular given several compiler optimizations to learn, 
%	and a primitive version of symbolic execution to learn
%	we show static models aren't 
%	nearly as comparable to a compiler as \dypro is.
\end{abstract}
\section{Introduction}
\label{sec:intro}

``Big Code'' emerged as a major line 
of research in the past decade. The idea 
is reusing the knowledge distilled from 
existing code repositories in an attempt to 
simplify the future development of software. 
Early methods applied NLP techniques to discover 
textual patterns existed in the source code
%off-the-shelf models developed for NLP tasks
~\cite{hindle2012naturalness,gupta2017deepfix,
	pu2016sk_p}; following approaches opted to 
learn the syntactic program embedding from the 
Abstract Syntax Trees (ASTs)~\cite{maddison2014structured,bielik2016phog,
	mou2016convolutional}. Although these pioneering 
efforts manage to transform programs in an 
amenable form to deep learning models, they 
only capture shallow, syntactic features and 
therefore are limited in what they can do. 
%Given the ultra-important role semantics play in 
%program analysis, learning representations for 
%program semantics has strong motivations. 
For example, a model that recognizes the reoccurring syntactic 
patterns may be sufficient for a code completion 
task; but to conquer more sophisticated and challenging 
problems in program synthesis or repair, thorough 
understanding and precise representations of program 
semantics can not be circumvented. 
Of late, a number of new deep learning 
architectures have been developed to specifically 
address this issue~\cite{
	wang2017dynamic,henkel2018code,
	allamanis2017learning,DeFreez}. Those works 
can be divided into two categories: dynamic and static. 
The former learns from program executions~\cite{
	wang2017dynamic,henkel2018code}. The latter dissects program 
semantics from source code. As an example,~\citet{allamanis2017learning} 
constructed a graph out of a program's AST. Subsequently they fed the 
graph to a Gated Graph Neural Network (GGNN)~\cite{li2015gated} 
for predicting variable misuse bugs in a method.

In this paper, we present a novel deep neural architecture, 
\dypro, that is capable of learning program semantics 
from execution traces. \dypro targets two 
major issues of the existing dynamic models. First, how to 
learn the dynamic semantics of a program as a whole rather than 
individual executions; second how to handle long execution traces that tend to 
hurt the generalization of underlying models. For the first 
challenge, we apply random testing, a powerful technique 
in software testing, to run a program 
with a large number of inputs, each of which will trigger a separate 
execution trace. \dypro then learns an embedding for each execution 
before compressing them into one vector that represents the 
semantics of the whole program. Regarding the second challenge, 
\dypro employs a bi-directional RNN to scan through an entire execution 
trace for filtering out the steps that are less essential to the 
trace. Our hope is \dypro could still learn a precise representation 
of the program semantics since the reduced traces would be likely to preserve 
the essence of the executions. More importantly, such reduction helps \dypro to 
handle longer traces more efficiently.

%\dypro is capable of compressing execution traces, 
%thus more resilient against program 
%yielding large runtime traces --- a well-known issue that 
%hinders dynamic analysis. 

We create a dataset to thoroughly evaluate how precise \dypro 
can capture the deep 
and rigid program semantics. The dataset consists of almost 
eighty-five thousand programs each of which solves one of ten coding 
problems. We pick the set of problems to cover a wide spectrum of difficulty 
levels ranging from entry-level programming exercises to 
challenging algorithmic questions frequently appearing on the 
coding interviews of major tech companies. The task is to 
classify programs in the dataset based on their semantics. 
For example, given a sorting routine, our goal is to test if 
models can differentiate among the algorithms that 
implement a sorting function (\eg \textit{Bubble Sort} and \textit{Insertion Sort} shown in 
Figure~\ref{fig:introexa}), an instance that in fact tricks 
all static models in our experiment. 
All programs in the dataset have been manually inspected and 
labeled before hand. Results show \dypro achieves significantly 
higher prediction accuracy than several prominent static models including 
GGNN, and TreeLSTM. We also 
found out that as the size of execution traces grows, \dypro 
suffers a smaller drop in prediction accuracy than Dynamic 
State Trace Neural Network (DSTNN)~\cite{wang2017dynamic}.
%while still beating all static models on the same set of programs. 

We conduct a second study to further 
examine if models can learn to detect deeper semantic 
properties of a program. Specifically, we choose loop invariants 
to evaluate all the above-mentioned deep neural architectures. 
Our intuition is recognition of loop invariants poses significant 
challenges to the model capacity. An effective model, at a bare minimum, 
needs to capture a precise representation of the program semantics; in 
addition it may also estimate the program properties derived from its 
semantics. Note our goal is not to invent models for generating loop 
invariants~\cite{NIPS2018_8001} but predicting loop invariants from a set 
of given properties. We use the classification accuracy as a means to 
gauge how precise and deep models have learnt the program semantics. 
Results show \dypro is far more accurate in predicting loop invariants 
than all static models. 
%the precision 
%of each model regarding learning representation of program semantics. 
%models learn to represent the program semantics. 
%learnt to represent the program semantics. Results show 
%evaluate 
%if models can learn a precise program representation so that they can 
%accurately classify the deep semantic properties like loop invariants. 
%To push the limits of deep neural networks' ability to dissect 
%program semantics, 
%%gain a deeper understanding of the results, 
%we conduct a further study to evaluate if models can learn to perform 
%several standard tasks of static analysis. In 
%particular, we choose three compiler optimizations: dead code elimination, 
%hoisting and induction variable elimination. We also test 
%if models can mimic the most primitive version of symbolic 
%execution. We find that all static models manage to learn the basic semantic features but fall short in 
%analyzing the deeper semantic properties such as recognizing the 
%loop invariant and examining the path feasibility. In contrast, 
%\dypro is highly precise in all tasks including some on which \dypro 
%outperforms all static models by a wide margin. 
Our findings indicate 
\dypro is the most precise deep neural network in learning 
representations of program semantics.

\vskip -0.08in
\begin{figure}[ht!]
	\begin{center}
		\begin{subfigure}[b]{0.45\textwidth}
			\lstset{style=mystyle}
\begin{lstlisting}[basicstyle=\linespread{.68}\fontsize{6}{10.8}\ttfamily\bfseries]
static int Difference(int[] a)
{
    //step 1: bubble sort the input array  
    int left = 0;
    int right = @a.Length - 1@;

    for (int i = @right@; i @> left@; i@--@) { 
        for (int j = @left@; j@ < i@; j@++@) { 
            if (a[j] > a[j + 1]) { 
                int tmp = a[j];
                a[j] = a[j + 1];
                a[j + 1] = tmp;
    }}}   
    //step 2: comput the largest difference 
    return a[a.Length-1] - a[0];
}
\end{lstlisting}		
		\end{subfigure}
%			\; %\quad
		\begin{subfigure}[b]{0.45\textwidth}
			\lstset{style=mystyle}
\begin{lstlisting}[basicstyle=\linespread{.68}\fontsize{6}{10.8}\ttfamily\bfseries]
static int Difference(int[] a)
{
    //step 1: insertion sort the input array
    int left = 0;
    int right = @a.Length@;

    for (int i = @left@; i @< right@; i@++@) {
        for (int j = @i-1@; j @>= left@; j@--@) { 
           if (a[j] > a[j + 1]) { 
              int tmp = a[j];
              a[j] = a[j + 1];
              a[j + 1] = tmp;
    }}}
    //step 2: same as the left 
    return a[a.Length-1] - a[0];
}
\end{lstlisting}	
		\end{subfigure}
	\end{center}
	\vskip -0.1in
	\caption{Two functions both compute the largest difference between two elements in a given array. 
		According to our evaluation, none of the static models recognize the semantic differences between 
		the two functions even though they apply distinct sorting routines: \textit{Bubble Sort} and \textit{Insertion Sort}. Code highlighted in shadow box are the only syntactic differences between the two functions.
	}
	\label{fig:introexa}    		
	\vskip -0.1in
\end{figure}

We make the following contributions: 
\begin{itemize}
	\item We design \dypro, a novel deep neural architecture capable of 
	learning dynamic program semantics from execution traces.
	
	\item We evaluate \dypro using a task of semantics classification. Results 
	show \dypro achieves the highest prediction accuracy among all competing models.
	
	\item We examine if models can learn to detect loop invariants. We find 
	\dypro is significantly more accurate than all static models.	
\end{itemize}
%\begin{itemize}
%	\item We design \dypro, a novel deep neural architecture capable of 
%	learning dynamic program semantics via execution traces.
%	
%	\item We evaluate \dypro using a task of semantic classification. Results 
%	show \dypro not only achieves the highest prediction accuracy among all 
%	competing models, but also scales better to long execution trace than 
%	DSTNN.
%
%	\item We investigate how \dypro's prediction accuracy vary with the code 
%	coverage achieved by executions in the training data.	
%\end{itemize}

%
%The remainder of this paper is organized as follows. Section~\ref{sec:back} gives the background. Section~\ref{sec:archi} 
%describes \dypro's architecture. Section~\ref{sec:exp} details the evaluation of 
%\dypro in semantic classification including the comparison against several other models. We survey related work in Section~\ref{sec:rela}. Finally we conclude in 
%Section~\ref{sec:con} with an outlook of future work.
\section{Background and Related Work}
\label{sec:back}

\vspace{-.1cm}
\paragraph{ML/DL for Source Code}
\citet{hindle2012naturalness} pioneered the field of 
learning language models from source code. A significant finding 
of theirs is programs, similar to natural language, are highly 
repetitive and predictable. With the rapid development of deep learning techniques and the increasing 
accessibility of large scale open source code repository, many works begin 
to apply deep neural networks to learn syntactic program representations
~\cite{gupta2017deepfix,pu2016sk_p,maddison2014structured,bielik2016phog,mou2016convolutional}. 
More recently, a new line of research emerges aiming to tackle the problem of 
learning semantic representation~\cite{wang2017dynamic,henkel2018code,allamanis2017learning,DeFreez}. 
%In this paper we present \dypro, a novel dynamic model that learns program 
%semantics from execution traces. Our evaluation shows \dypro is more accurate in understanding 
%and analyzing program semantics than several prominent static models.

\vspace{-.1cm}
\paragraph{Execution-Based Program Representation}
Neural program interpreter~\cite{reed2015neural} is the first 
deep neural network that deal with programs in the form of 
execution trace. The idea is to learn to synthesize a sequence 
of low-level primitive operations for solving a high-level task 
such as addition and sorting. Later Cai~\etal propose an improvement~\cite{cai2017making} 
by addressing the NPI's generalization issues. In particular they replace loops with 
recursions that used to synthesize the low-level operations. Unlike 
the prior attempts which still rely on the syntactical representation 
of execution traces,~\citet{wang2017dynamic} 
%proposes two 
%schemes that 
encode each step of the execution trace as a snapshot 
of the memory. Their intuition is such processing, widely regarded as 
dynamic analysis in the field of program analysis, leads to a direct and 
concrete expression of program semantics. As a result, models 
can work with the program semantics while totally detaching 
themselves from the program syntax. Furthermore, 
~\citet{wang2017dynamic} deals with real world programs, albeit mostly 
entry-level programming exercises, whereas the others work with 
artificial programs written in low-level language.

In this paper, we choose state trace encoding scheme over variable 
trace encoding scheme for better precision. State trace encoding 
scheme also enables a far more efficient 
implementation in DSTNN that is almost as accurate as Dependency Enforcement 
Network according to Wang et al.'s evaluation~\cite{wang2017dynamic}.
We address two major weaknesses of DSTNN. First 
DSTNN learns semantic representations for individual executions 
rather than a entire program; second DSTNN has difficulty in handling 
programs yielding long execution traces. 

\vspace{-.1cm}
\paragraph{Random Testing}
Random testing is a software testing technique where programs are 
tested by generating random, independent inputs. Results of the output 
are compared against software specifications to verify the test output 
is pass or fail. Alternatively, random testing can also be used to 
catch the exceptions of the language (\ie crashes) which means if an 
exception arises during testing execution then it means there is a 
fault in the program.

Random testing is practical, effective that is shown to be able to 
achieve good coverage in a variety of testing 
domains~\cite{5010257,Patel:2018}. As a result, it's one of the most powerful and 
widely adopted testing technique in software testing.

\section{The Model Architecture of \dypro}
\label{sec:archi}
%In this section, we start with an overview of the 
%overall architecture of \dypro. Next we present a 
%formalization of  \dypro's workflow.
%The 
%design decision are made solely to address the shortcomings  
%of the models proposed in~\cite{wang2017dynamic}. 
%In particular, none of their three architectures dealt with the 
%scalability issues. That is when programs are being executed, 
%the traces can grow arbitrarily long, making it difficult for RNNs 
%to generalize. Furthermore, strictly speaking their models do not 
%learn the program semantics, instead, they are trained on individual 
%execution traces.   
%
%In this paper, we choose to refine the state trace model proposed in
%\cite{wang2017dynamic} because it requires lower engineering effort while 
%only being marginally outperformed by the dependency enforcement network.
%Figure~\ref{fig:archi} depicts the overall architecture of \dypro. 
%We first give an overview of the overall architecture of \dypro 
%(Figure~\ref{fig:archi}) followed by a formalization of the model.
%As annotated 
%in the following diagram, 
%We split the entire network into five layers and 
%give a detailed explanation for each layer respectively.

\begin{figure*}[t]
	\vskip 0.2in
	\begin{center}
		\centerline{\includegraphics[width=1\columnwidth]{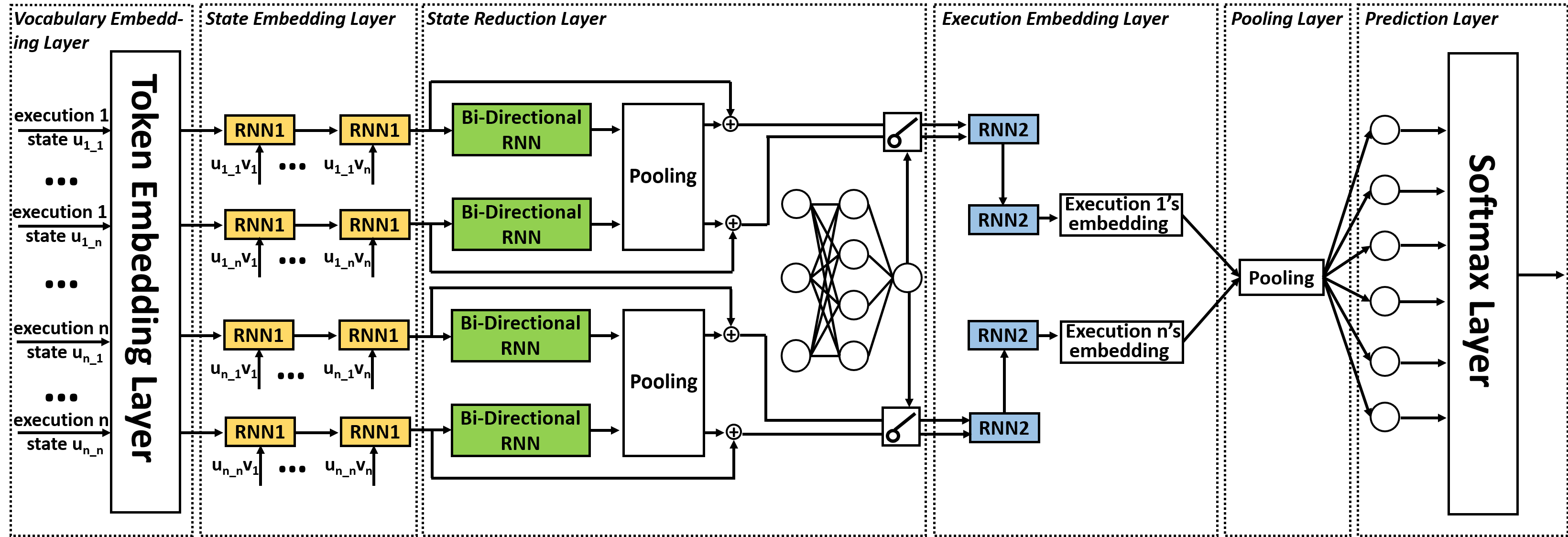}}
		\caption{Architecture of \dypro.}
		\label{fig:archi}
	\end{center}
	\vskip -0.2in
\end{figure*}

%\subsection{Overview}
In a general description, \dypro consists of two recurrent 
neural networks (RNN) and another bi-directional RNN. Given 
the execution traces triggered by random testing, we encode 
each trace into a sequence of program states. Specifically, a 
program state is a tuple of values based on the variable 
valuations (\ie the state of the memory). Each program state is 
created due to a memory update issued by a statement/instruction. 
By collecting the history of memory updates occurred during an 
execution, one can convert the trace into a sequence of program states. 
%Upon receiving 
%the inputs, \dypro uses the first RNN (\ie 
%RNN1 in Figure~\ref{fig:archi}) to encode each program 
%state into an embedding vector. Next \dypro employs a bi-directional 
%RNN to scan through the entire sequence of state embeddings 
%for picking the states that capture the essence of the 
%execution. In other words, state embeddings that are not 
%selected will be pruned away. After we feed the resulting 
%state embeddings as a sequence to the second RNN (\ie 
%RNN2 in Figure~\ref{fig:archi}), we extract its final 
%state to represent an execution. Then we perform 
%max pooling across embeddings that are learnt 
%from executions of the same program to obtain the program 
%embedding. Finally, we add a one layer softmax regression to 
%make predictions. 
%\subsection{Formalization}
Formally, given a program $\mathit{P}$, and its variable set 
$\mathit{V}$ ($\mathit{v}_1$, $\mathit{v}_2$,..,$\mathit{v}_n$ $\in$ $\mathit{V}$), 
%we represent an 
%execution of $P$ as a sequence of program states, each 
%of which is a tuple of variable values. 
%%created while running $\mathit{P}$ with an input. 
%For example, 
we encode $\mathit{S}_{t\_e}$, the 
$t$-th program state in $e$-th execution of $\mathit{P}$, 
to be $(x_{t\_e\_v_{1}}$,$x_{t\_e\_v_{2}}$,..,$x_{t\_e\_v_{n}})$, 
where $x_{t\_e\_v_{n}}$ is the value of variable $\mathit{v}_n$ 
in $\mathit{S}_{t\_e}$. Using this notation, we explain \dypro's working pipeline.

%\paragraph{Embedding Tokens}  First, we embed each 
%variable value in each program state into a vector.

\paragraph{Embedding Program States}  Each program state will 
be embedded by RNN1 into a vector. Take $\mathit{S}_{t\_e}$ 
for example, we feed the variable values (\ie $x_{t\_e\_v_{1}}$ 
to $x_{t\_e\_v_{n}}$) as a sequence to RNN1 and extract its 
final hidden state $h_{t\_e}$ (Equation (1)) as the state embedding vector of $\mathit{S}_{t\_e}$. 
$h_{t\_e\_v_{1}}$,..,$h_{t\_e}$ $\in$ $\mathbb{R}^{k_{1}}$ where $k_{1}$ denotes the size of the hidden layer of RNN1.
Note that we do not assume the order in which variables values are encoded 
by RNN1 for each program state but rather maintain a consistent order 
throughout all states for a given trace. 

\vskip -0.2in
\begin{small}
\begin{align*}	
h_{t\_e\_v_{1}}\!&=\!\text{RNN1} (h_{t\_e\_v_{0}}, x_{t\_e\_v_{1}}) \!\!\!\!& h_{t\_e\_v_{2}}\!&=\!\text{RNN1} (h_{t\_e\_v_{1}}, x_{t\_e\_v_{2}}) \!\!\!\!\!\!\!& ... h_{t\_e}\!&=\!\text{RNN1} (h_{t\_e\_v_{n-1}}, x_{t\_e\_v_{n}})\,\text{(1)}
\end{align*}
\end{small}

\vspace{-.4cm}
\paragraph{Improving \dypro's Scalability}  \textit{State reduction layer is dedicated to 
enhancing \dypro's resilience against programs yielding long 
execution trace}. We describe its inner workings as follow. After obtaining all the state embeddings, we feed them into a 
bi-directional RNN which aims to filter out those that are 
less essential to the execution. Formally, let 
$h_{1\_e}$ to $h_{m\_e}$ denote the whole 
sequence of state embedding vectors for $e$-th execution of 
$\mathit{P}$, we register the hidden vectors $\overrightarrow{H_{1\_e}}$ 
to $\overrightarrow{H_{m\_e}}$ of the forward one in the bi-directional RNN as 
defined in Equation (2). Similarly we also extract 
$\overleftarrow{H_{1\_e}}$ to $\overleftarrow{H_{m\_e}}$ 
out of the backward RNN defined in Equation (3). Next for each 
program state, we define two context vectors to represent 
its prior and subsequent execution using 
$\overrightarrow{H_{1\_e}}$ to $\overrightarrow{H_{m\_e}}$ 
and $\overleftarrow{H_{1\_e}}$ to $\overleftarrow{H_{m\_e}}$. 
For example, given embedding vector $h_{t\_e}$, we define the two context vectors 
$C_{f}(h_{t\_e})$ and $C_{b}(h_{t\_e})$ in Equation (4). Finally 
we concatenate the context vectors with the state embedding vector 
to determine if a program state should be retained or discarded. Specifically, 
we train a Multi-Layer Perceptron\footnote{with one single sigmoid output neuron.} (MLP) to produce a mask (Equation (5) where $\oplus$ denotes
vector concatenation). Equation (6) applies the mask on the program states; 
$\odot$ denotes element-wise 
matrix multiplication assuming the broadcasting behavior. 
Our intention is to equip \dypro 
with the capability of selecting program states that are most 
essential to the execution while discarding others that somewhat recoverable. 
$\overleftarrow{H_{1\_e}}$,..,$\overleftarrow{H_{m\_e}}$; 
$\overrightarrow{H_{1\_e}}$,..,$\overrightarrow{H_{m\_e}}$; 
$C_{f}(h_{t\_e})$; and $C_{b}(h_{t\_e})$ $\in$ $\mathbb{R}^{k^{'}}$ 
where $k^{'}$ denotes the size of the hidden layer of the bi-directional RNN. 
$\mathit{M}(h_{t\_e})$ $\in$ (0,1). 

\vskip -0.2in
\begin{small}
\begin{align*}	
\overrightarrow{H_{1\_e}}\!&=\!\text{FR} (\overrightarrow{H_{0\_e}}, h_{1\_e}) \!\!\!\!& \overrightarrow{H_{2\_e}}\!&=\!\text{FR} (\overrightarrow{H_{1\_e}}, h_{2\_e}) \!\!\!\!\!\!\!& ... \;\;\;\overrightarrow{H_{m\_e}}\!&=\!\text{FR} (\overrightarrow{H_{m-1\_e}}, h_{m\_e})\;\;\;&\text{(2)}\\
\overleftarrow{H_{1\_e}}\!&=\!\text{BR} (\overleftarrow{H_{0\_e}}, h_{m\_e}) \!\!\!\!& \overleftarrow{H_{2\_e}}\!&=\!\text{BR} (\overleftarrow{H_{1\_e}}, h_{m-1\_e}) \!\!\!\!\!\!\!& ... \;\;\;\overleftarrow{H_{m\_e}}\!&=\!\text{BR} (\overleftarrow{H_{m-1\_e}}, h_{1\_e})\;\;\;&\text{(3)}
\end{align*}
\end{small}

\vspace{-.48cm}
\vskip -0.2in
\begin{small}
\begin{align*}	
C_{f}(h_{t\_e})\!&=\!\text{pooling}(\overrightarrow{H_{1\_e}}, \overrightarrow{H_{2\_e}},..,\overrightarrow{H_{t-1\_e}})& C_{b}(h_{t\_e})\!&=\!\text{pooling}(\overleftarrow{H_{1\_e}}, \overleftarrow{H_{2\_e}},..,\overleftarrow{H_{m-t\_e}})\;\;\,\,\,\,\text{(4)}
\end{align*}
\end{small}
\vspace{-.55cm}
\vskip -0.2in
\begin{small}
	\begin{align*}
	\;\;\;\;\;\;\;\;\;\;\;\;\;\;\;\mathit{M}(h_{t\_e})\!&=\!\text{MLP}(C_{f}(h_{t\_e})\oplus C_{b}(h_{t\_e})\oplus h_{t\_e}) \:\!\;\;\;\;\;\;\;\;\;\;\;\;\;\;\;\;\;\;\;\;\;\;\;\;\;\;\;\;\;\;\;\;\;\;\;\;\;\;\;\;\text{(5)} \\
	\;\;\;\;\;\;\;\;\;\;\;\;\;\;\;[h_{1\_e}^{'};\!h_{2\_e}^{'};\!..;\!h_{m\_e}^{'}]\!&=\!	
	[h_{1\_e};\!h_{2\_e};\!..;\!h_{m\_e}]
	\odot
	[\mathit{M}(h_{1\_e});\!\mathit{M}(h_{2\_e});\!..;\!\mathit{M}(h_{m\_e})]\;\;\;\;\;\;\;\;\;\;\;\;\;\text{(6)}
	\end{align*}
\end{small}
%\vspace{-.7cm}
%\vskip -0.2in
%\begin{small}
%	\begin{align*}
%	\;\;\;\;\;\;\;\;\;\;\;\;\;\;\;
%	[h_{1\_e}^{'};\!h_{2\_e}^{'};\!..;\!h_{m\_e}^{'}]\!=\!	
%	[h_{1\_e};\!h_{2\_e};\!..;\!h_{m\_e}]
%	\odot
%	[\mathit{M}(h_{1\_e});\!\mathit{M}(h_{2\_e});\!..;\!\mathit{M}(h_{m\_e})]\;\;\;\;\;\;\;\;\;\;\;\;\;\text{(6)} \end{align*}
%\end{small}

\vspace{-.4cm}
\paragraph{Embedding Executions} Given $h_{1\_e}^{'}$, 
$h_{2\_e}^{'}$,...,$h_{m\_e}^{'}$, the state embeddings 
preserved by the previous layer. Equation (7) computes 
$h_{e}^{''}$, the embedding vector that represents the 
whole execution. $h_{1\_e}^{''}$, $h_{2\_e}^{''}$,..,$h_{e}^{''} \in \mathbb{R}^{k_{2}}$ 
where $k_{2}$ denotes the size of the hidden layer of RNN2.
\vskip -0.2in
\begin{small}
	\begin{align*}	
	h_{1\_e}^{''}\!&=\!\text{RNN2} (h_{0\_e}^{''}, h_{1\_e}^{'}) & 
	h_{2\_e}^{''}\!&=\!\text{RNN2} (h_{1\_e}^{''}, h_{2\_e}^{'}) & 
	\!\!\!\!\!\!...\;\;\;\;\; 
	h_{e}^{''}\!&=\!\text{RNN2} (h_{m-1\_e}^{''}, h_{m\_e}^{'})\,\;\;\;\;\;\;\;\text{(7)}
	\end{align*}
\end{small}
\vspace{-.6cm}
\vskip -0.2in
\begin{small}
	\begin{align*}
	\;\;\!
	h_{\mathit{P}} &= \text{max\_pooling} (h_{1}^{''}, h_{2}^{''},..,h_{q}^{''}) \;\;\;\;\;\;\;\;\;\;\;\text{(8)} \;\;\;\;\;\;\;\;\;\;\;\;\;\;\;\;\;\;\;\;\;
	L = \mathit{H} + \sum_{e=1}^{q} \sum_{t=1}^{m} \mathit{M}(h_{t\_e}) \;\;\;\;\;\;\;\;\;\;\; \text{(9)}
	\end{align*}
\end{small}

%
%
%\begin{minipage}{0.42\textwidth}
%\begin{align*}	
%h_{t\_e\_v_{n}}^{'} &= \text{RNN2} (h_{t-1\_e\_v_{n}}^{'}, h_{t\_e\_v_{n}}) \;\;\;\;\;\;\;\;\,\,\text{(7a)}\\
%h_{t+1\_e\_v_{n}}^{'}     &= \text{RNN2} (h_{t\_e\_v_{n}}^{'}, h_{t+1\_e\_v_{n}}) \;\;\;\;\;\;\;\;\,\,\text{(7b)}\\
%\,\,&\,\:\!...\\
%h_{e\_v_{n}}^{'} &= \text{RNN2} (h_{t+q-1\_e\_v_{n}}^{'}, h_{t+q\_e\_v_{n}})\,\,\,\text{(7c)} 
%\end{align*}
%\end{minipage}
%\hspace{-.15cm}
%\begin{minipage}{0.496\textwidth}
%	\begin{align*}	
%	h_{\mathit{P}} &= \text{max\_pooling} (h_{1\_v_{n}}^{'}, h_{2\_v_{n}}^{'},..,h_{e\_v_{n}}^{'}) \tag{8}\\[2pt]
%%	\sigma &= \text{softmax} (\text{MLP}(h_{\mathit{P}})) \tag{9} \\
%\\
%	l &= \sum_{e=1}^{n_{e}} \sum_{t=1}^{n_{t}} \mathit{M}(h_{t\_e\_v_{n}}) \tag{9}
%	\end{align*}
%\end{minipage}
%
\vspace{-.4cm}
\paragraph{Embedding Programs} \textit{Pooling layer is solely responsible for distilling the program representation 
from all execution embeddings}. Let $h_{1}^{''}$, $h_{2}^{''}$,..,$h_{q}^{''}$ denote the 
embedding vectors for all executions of $\mathit{P}$. Equation (8) computes 
the embedding vector of program $\mathit{P}$. $h_{\mathit{P}} \in \mathbb{R}^{k_{2}}$.

%Finally we add a dense layer 
%followed by the softmax operation to make the predictions. 
%$h_{t\_e\_v_{n}}^{'}$,..,$h_{e\_v_{n}}^{'}\!$; 
%$h_{1\_v_{n}}^{'}$,..,$h_{e\_v_{n}}^{'}$ and $h_{\mathit{P}}$ 
%$\in$ $\mathbb{R}^{k_{2}}$ 
%where $k_{2}$ denotes the size of the hidden layer of RNN2.

%\paragraph{\textbf{Execution Embedding Layer}}
%After reducing the program states, we use a third RNN to 
%encode the entire execution into an embedding vector.
%
%\paragraph{\textbf{Pooling Layer}}
%We max-pool across all execution embeddings to  
%obtain the program embedding.
%
%\paragraph{\textbf{Prediction Layer}}
%A dense layer followed by the softmax layer to make 
%the predictions.

\paragraph{Designing Loss Function} The network is trained to minimize the cross-entropy loss 
on a softmax over the semantic labels (denoted by $\mathit{H}$ in Equation (9))
along with the sum of $\mathit{M}(h_{t\_e})$ \wrt 
all program states among all executions of $\mathit{P}$. At a high-level, we force \dypro to cut 
as many states from each trace as possible 
provided that it can still maintain a high prediction accuracy.

\section{Evaluation}
\label{sec:exp}

We present the evaluation of \dypro which begins with 
semantics classification followed by the detection of loop 
invariants.

\subsection{Semantics Classification}
\label{subsec:coset}
%The goal of the experiment is to measure how precise the 
%deep neural architectures can learn the program semantics.

\paragraph{\textbf{Dataset}}
The dataset consists of 84,165 programs in total. 
They are obtained from a popular online coding platform. 
Programs were written in several different languages: Java, C\# and Python. 
All programs solve a particular coding problem. We hand picked the 
problems to ensure the diversity of the programs in the dataset. 
Specifically, it contains introductory programming exercises for 
beginners, coding puzzles that exhibit considerable algorithmic 
complexity and challenging problems frequently appearing on coding interviews. 
The dataset was manually analyzed and labeled. 
The work was done by fourteen PhD students and exchange scholars at University of 
California, Davis. To reduce the labeling error, we distributed programs solving 
the same coding problem mostly to a single person and had 
them cross check the results for validation. The whole process 
took more than three months to complete. Participants came from different research 
backgrounds such as programming language, database, security, graphics, machine learning, 
\etc All of them were interviewed and tested for their knowledge on program 
semantics. The labeling is on the basis of operational semantics (\eg bubble sort, insertion sort, 
merge sort, \etc, for a sorting routine). We allow certain kind of variations to keep the total 
number of labels manageable. For example, we ignore local variables allocated for temporary 
storage, the iterative style of looping or recursion, the sort order: descending or ascending. 
Readers are invited to consult the supplemental material for the descriptions of all coding problems and the list of all labels. 

\begin{figure*}[t!]
	\begin{subfigure}[b]{0.325\textwidth}
		\begin{center}
			\centerline{\includegraphics[width=\columnwidth]{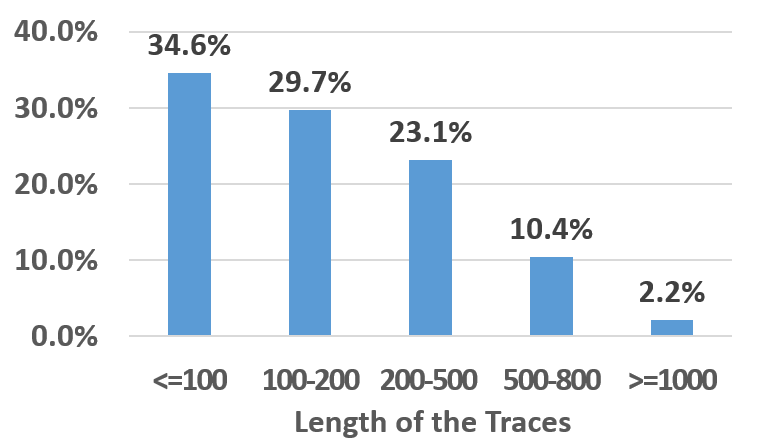}}
			\caption{Traces distribution.}
			\label{fig:ins}
		\end{center}
	\end{subfigure}
	\begin{subfigure}[b]{0.325\textwidth}
		\begin{center}
			\centerline{\includegraphics[width=\columnwidth]{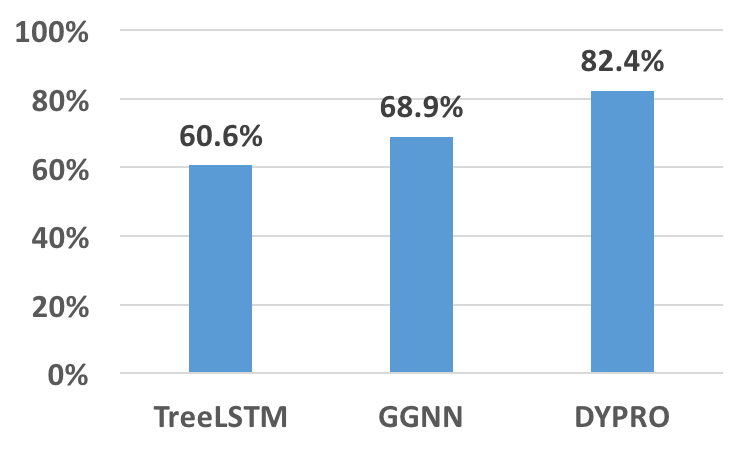}}
			\caption{Prediction (Accuracy).}
			\label{fig:acc}
		\end{center}
	\end{subfigure}
%	\; %\quad
	\begin{subfigure}[b]{0.325\textwidth}		
		\begin{center}
			\centerline{\includegraphics[width=\columnwidth]{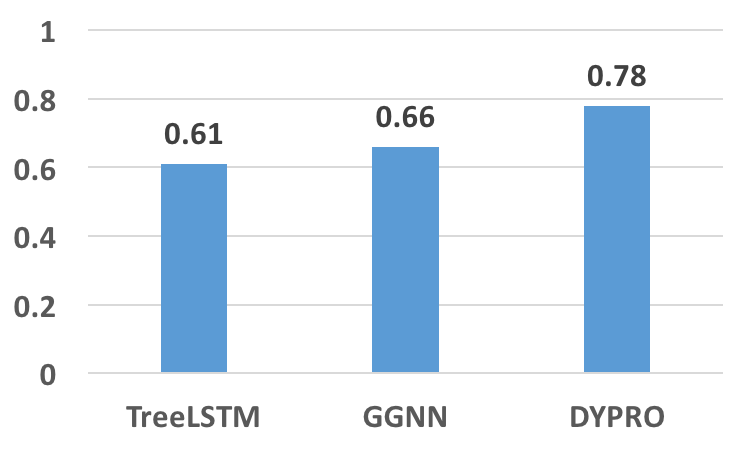}}
			\caption{Prediction (F1 Score).}
			\label{fig:accF1}
		\end{center}
	\end{subfigure}
	\begin{subfigure}[b]{0.325\textwidth}		
		\begin{center}
			\centerline{\includegraphics[width=\columnwidth]{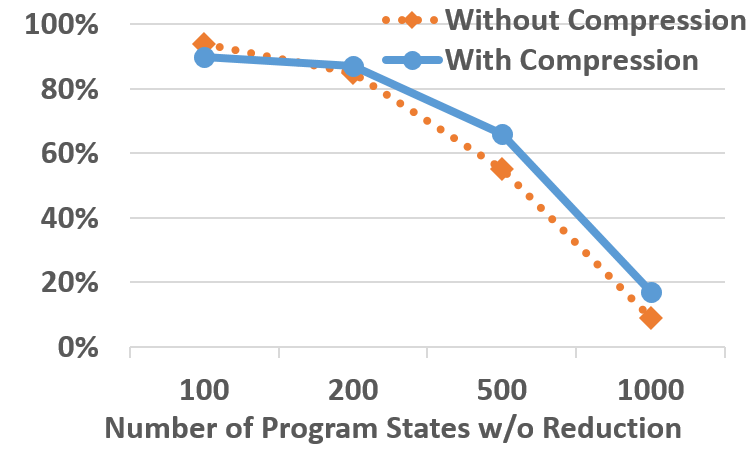}}
			\caption{Scalability (Accuracy).}
			\label{fig:sca}
		\end{center}
	\end{subfigure}
%	\; %\quad
	\begin{subfigure}[b]{0.325\textwidth}
		\begin{center}
			\centerline{\includegraphics[width=\columnwidth]{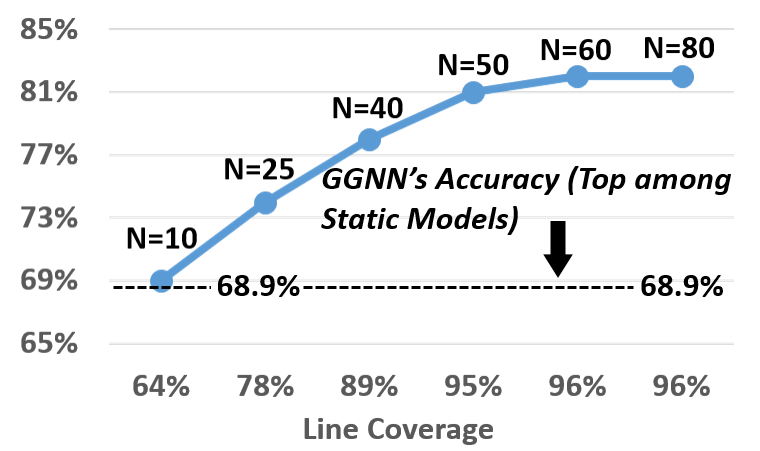}}
			\caption{Coverage (Accuracy).}
			\label{fig:cov}
		\end{center}
	\end{subfigure}
	\begin{subfigure}[b]{0.325\textwidth}
		\begin{center}
			\centerline{\includegraphics[width=\columnwidth]{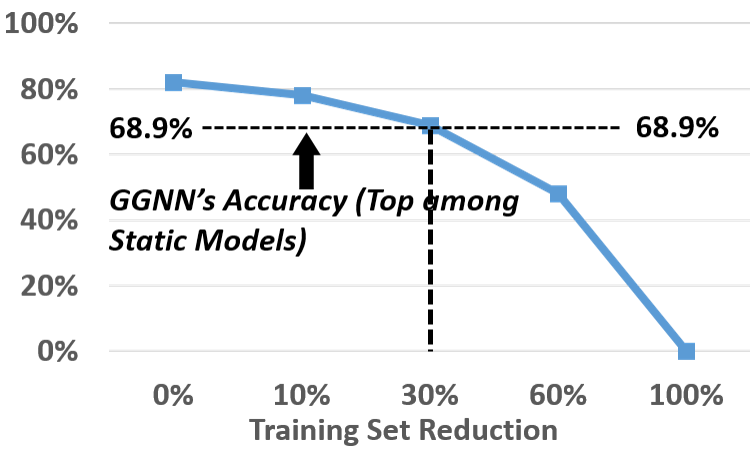}}
			\caption{Reduction (Accuracy).}
			\label{fig:redu}
		\end{center}
    \end{subfigure}
%	\; %\quad
	\begin{subfigure}[b]{0.325\textwidth}		
		\begin{center}
			\centerline{\includegraphics[width=\columnwidth]{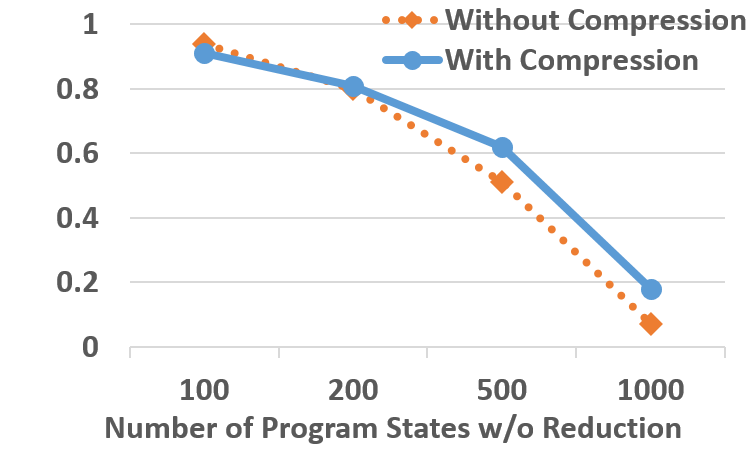}}
			\caption{Scalability (F1 Score).}
			\label{fig:scaF1}
		\end{center}
	\end{subfigure}
%	\; %\quad
	\begin{subfigure}[b]{0.325\textwidth}
		\begin{center}
			\centerline{\includegraphics[width=\columnwidth]{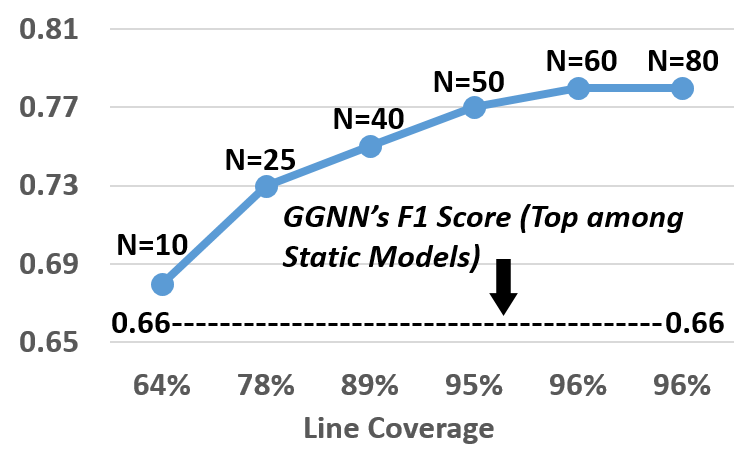}}
			\caption{Coverage (F1 Score).}
			\label{fig:covF1}
		\end{center}
	\end{subfigure}
	\begin{subfigure}[b]{0.325\textwidth}
		\begin{center}
			\centerline{\includegraphics[width=\columnwidth]{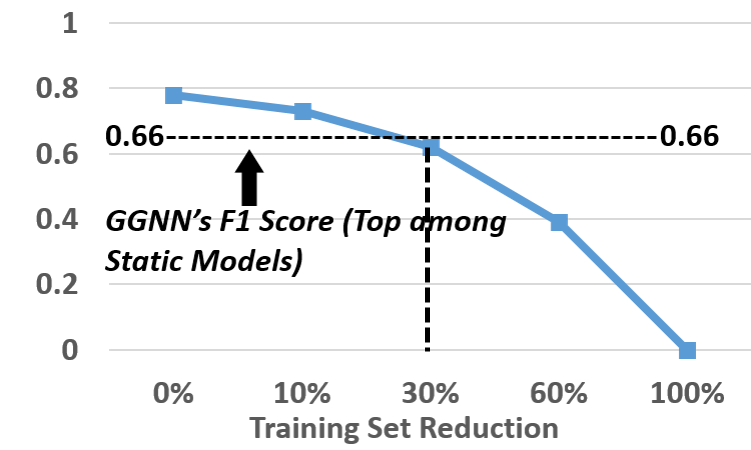}}
			\caption{Reduction (F1 Score).}
			\label{fig:reduf1}
		\end{center}
	\end{subfigure}
	
	\caption{Evaluation Results.}
	\label{fig:ss}
	\vskip -0.1in
\end{figure*}

To seek random inputs for producing the execution traces, we separate 100 
programs for each coding problem to apply random 
testing. In particular we first generate $\mathit{N}$ different 
inputs to execute each program and then select top $\mathit{N}$ 
(out of $100*\mathit{N}$) inputs that achieve the highest line 
coverage to be the test cases for all programs. We remove 
programs that fail to pass all the test cases (\ie crashes or 
incorrect outputs) from the dataset. In the end we are left with 72,376 
programs which we split into a training set containing 52,376 programs, 
a validation set of 10,000 programs and a test set of the remaining 
10,000 programs (Table~\ref{Table:data}). Figure~\ref{fig:ins} depicts the cumulative 
percent distribution of all traces by length. We calculate the 
average length of an execution trace to be 218 (median=192).

%100: 34.6.      0.9       3425.4      1106078.96
%100-200: 29.7.  0.85      2776.9      53679071.165
%200-500: 23.1.  0.7.      1701.7      5538386.7
%500-1000:10.4.  0.5.      627.        6649260.8
%1000: 2.2.      0.2.      48.4 

%In the end, we have 53 labels in total and 
%on average more than 1,000 programs for each label. We split the whole 
%dataset into a training set containing 55,153 programs, a validation set of 11,000 programs and 
%a test set of the remaining 11,000 programs (Table~\ref{Table:data}).  
%also 
%describes the average LOC (\ie lines of code) and 
%SOT (\ie size of traces) for programs solving the same problem. 
\begin{table}[htbp!]
	\begin{center}
		\begin{adjustbox}{max width=.8\textwidth}
			\begin{tabular}{c | c | c | c } %| c | c} 
				\hline
				\textbf{Benchmarks} 
				& \textbf{Training}
				& \textbf{Validation}
				& \textbf{Testing}
				%				& \textbf{LOC}
				%				& \textbf{SOT}
				\\
				%				\hline
				
				%				\hline 
				%				Array Average &4,644  &1,000  &1,666 \\						
				%				\hline 
				%				Array k\textsuperscript{th} Largest &3,639  &1,000  &1,666 \\										
				\hline		
				Print Chessboard  &7,415  &1,000  &1,000 \\ %for benchmark 11,318	
				\hline
				Find Array Max Difference &5,821  &1,000 &1,000 \\ %for benchmark 8,162	   									
				\hline
				Check Matching Parenthesis  &3,269 &1,000  &1,000 \\ %6,249
				\hline
				Reverse a String  &5,946  &1,000  &1,000 \\ %9,100										
				\hline		
				Sum of Two Numbers  &6,635  &1,000  &1,000 \\							
				\hline		
				Find Extra Character  &5,020  &1,000  &1,000 \\ 			
				\hline		
				Maximal Square  &4,836  &1,000  &1,000 \\ 			
				\hline		
				Maximal Product Subarray &5,174  &1,000  &1,000 \\ 				
				\hline		
				Longest Palindrome &5,273  &1,000  &1,000 \\ 				
				\hline		
				Trapping Rain Water &2,987  &1,000  &1,000 \\							
				\hline
				\hline
				Total &52,376 &10,000  &10,000 \\ %&27 &108\\										
				\hline
			\end{tabular}
		\end{adjustbox}
	\end{center}
	\caption{Dataset used in semantics classification.}
	\label{Table:data}
\end{table}

\paragraph{\textbf{Prediction Task}}
Similar to the image classification setting, models are required to predict the 
category a program falls into based on its semantics. In 
other words, not only do models need to classify which coding problem 
a program attempts to solve but also how the program solves it. We adopt 
prediction accuracy and F1 score as the evaluation metrics.

\paragraph{\textbf{Evaluation Subjects}}
Apart from \dypro, We select GGNN, arguably the state-of-the-art 
deep neural network for learning source code. We also 
include TreeLSTM~\cite{TreeLSTM}, one of the mostly applied  
deep neural networks for learning data structure of trees, 
in our case the ASTs of programs.

\paragraph{Implementation}
We use Roslyn, IronPython and Eclipse JDT for parsing 
programs written in C\#, Python and Java. All models are implemented in 
Tensorflow. Before training, we have unified as many 
hyperparameters as possible across all models such as the number of recurrent layers: 1; the 
number of hidden unit in the recurrent layer: 100; the embedding 
dimensions for each input token: 100; the optimizer: the Adam algorithm; 
the maximum value for normalizing the gradient when clipping: 0.9, 
\etc We use two Red Hat Linux servers each of which host two 
Tesla V100 GPUs (of 32GB GPU memory). Training \dypro took the longest: 
approximately 48 hours in total while all other models finished 
within three hours.

\paragraph{Results}
Figure~\ref{fig:acc} shows the prediction accuracy of 
all models. \dypro leads the pack by almost 15\%. 
%All models are very accurate 
%in classifying programs based on their functionality, 
%which confirms the hypothesis they are capable of 
%learning syntactic features. However, all models suffer notable 
%drops in semantic classification even 
%though 
%\dypro remains to be the most accurate beating GGNN who 
%comes next by almost fifteen percent. 
TreeLSTM is barely above 60\%. In terms of the F1 score, \dypro also achieves 
better results than all others (Figure~\ref{fig:accF1}). Regarding 
scalability, Figure~\ref{fig:sca} (\resp \ref{fig:scaF1}) depicts how \dypro 
's prediction accuracy (\resp F1 score) vary with the size of execution 
traces. As a baseline for comparison, we re-implemented \dypro without 
the state reduction layer. Results show as 
the number of program states reaches two hundred, \dypro starts to 
outperform the baseline configuration, especially the gap grows to be 
approximately 10\% in accuracy (\resp 0.1 in F1 score) for traces of five 
hundred program states. To measure \dypro's capability of compressing execution traces, 
we find on average \dypro discarded 27\% (median=26\%) 
of the program states for each execution trace among all testing programs. 
Finally we investigate the influence of code coverage on \dypro's prediction. 
According to Figure~\ref{fig:cov} and \ref{fig:covF1}, when $\mathit{N}$ 
%(\ie the number of test cases for each program in the dataset) 
is initialized to 10, we begin with on average 64\% line coverage (among all 
programs in the test set), where \dypro already outperforms all 
static models in both prediction accuracies and F1 score, albeit by 
a small margin. As $\mathit{N}$ continues to grow, line coverage monotonically increases. So 
does \dypro's prediction accuracy and F1 score. After $\mathit{N}\!$=60, additional 
test inputs produce no further improvement beyond 96\% line coverage. 
%Note even at the lowest coverage point, \dypro still outperforms all 
%static models in both prediction accuracies and F1 score, albeit by a small margin. 
Overall we demonstrate random testing lays a good 
foundation for \dypro.

\paragraph{Analysis}
To investigate the cause of inaccuracies, we look 
into the misclassifications static models produced. In general, 
we see two common classes of errors.
%that cover a large portion of 
%the misclassifications static models make
First as briefly 
mentioned in Section~\ref{sec:intro}, when given syntactically 
similar programs in Figure~\ref{fig:introexa}, static models 
struggle to differentiate their semantic differences. This 
type of mistakes indicates the insufficiency of the underlying 
program representation \wrt the provided model capacity. 
%Apart from 
%the programs in Figure~\ref{fig:introexa}, we show 
%another example in Figure~\ref{fig:mis} that are 
%representative to a large portion of errors we have 
%seen static models make. First to explain the 
%misclassifications for programs in Figure~\ref{fig:introexa}, 
%
%Although static model are shown 
%to be capable of learning semantics to some degree, they 
%still have trouble in separating program syntax while 
%extracting program semantics. Given the two programs depicted in 
%Figure~\ref{fig:sssd}, none of the static models produces 
%the correct classification even though the two sorting 
%routines (\textit{Bubble} sort on the left and \textit{Insertion} 
%sort on the right) express clearly different semantics. 
%
%As introduced in Section~\ref{sec:intro}, 
%
%
%The reason of this mistake is apparent: two program have 
%very similar ASTs. Even taking into account the additional 
%edges Allamanis~\etal introduced for GGNN or the path information 
%APNN utilizes, there weren't enough evidence for any static model 
%to differentiate the two programs. 
Similarly, when programs are written in vastly different syntax, 
%like what's depicted in Figure~\ref{fig:sdss}, 
static models have difficulty in recognizing the 
same semantics programs denote.\footnote{We have provided several 
examples in the supplemental material.} Our findings indicate that static 
models still largely generalize at the level of program syntax. Although 
certain semantic features can be learnt, their 
generalization will result in imprecise modeling of program 
semantics. In contrast, \dypro correctly classified all example programs, 
%in Figure~\ref{fig:introexa} and~\ref{fig:sdss},
especially the syntactic variations that hindered static models are automatically 
canonicalized by the executions. This observation gives rise to a 
hypothesis---that is, due to the unification of syntactic discrepancies 
by the runtime execution, we expect \dypro to 
present a similar capacity with the static models despite learning from a 
smaller set of training data. To confirm this hypothesis, we conduct another experiment 
in which we randomly remove programs from the original training set. As depicted in 
Figure~\ref{fig:redu} and Figure~\ref{fig:reduf1}, using approximately 70\% of the 
training data, \dypro is almost as accurate as the static models. 

However, \dypro can also 
be inaccurate at times for the following reasons. Although 
reducing program states benefits \dypro overall, it also causes 
\dypro to misclassify which it otherwise wouldn't. By simply 
removing the state reduction layer, \dypro remedies 
more than 10\% of the misclassifications. On the other 
hand, 31\% of the misclassifications are due to programs 
yielding long execution traces (\ie more than 
eight hundred program states) indicating \dypro still has difficulties 
in generalizing longer traces. This phenomenon necessitates the split of 
the problem into smaller ones which can be solved with separate tactics. 
In particular, when given shorter traces, 
\dypro may chose to skip the state reduction layer and 
take the trace in its entirety to prioritize the precision 
of the learning. On the contrary, for much longer traces, 
\dypro can be more aggressive in trace reduction 
since trading precision for scalability is generally worthwhile 
as shown in this experiment.

\subsection{Detection of Loop Invariants}
\label{subsec:ana}

As a more challenging task, we evaluate if models 
can recognize loop invariants --- properties of 
loops that are true before (and after) each iteration --- from 
a set of program expressions. Worth noting in this paper we 
do not formally infer loop invariants which requires extra 
functionality such as search or logic deduction. Instead, our rationale is 
given deeper program properties like loop invariants, a capable 
model not only would understand the program semantics, 
but may also ``infer'' the properties determined by the 
semantics. In other words, a model with high prediction 
accuracy is a testament to its capability of learning 
deep and precise representation of program semantics.

\paragraph{\textbf{Data Preparation}}
In order to reuse the dataset introduced in Section~\ref{subsec:coset}, 
we need to find loop invariants as our training labels. Here is our methodology. 
For each program in the dataset, first we use Daikon~\cite{ernst2007daikon} 
to propose the likely loop invariants in each loop. Since Daikon’s output forms a 
specification from the view of a client for each procedure, it 
does not produce invariants for local variables, neither does it 
within a procedure. To address the issue, we convert 
all local variables within a loop to be members of the class; 
in addition we insert a dummy procedure both at the beginning and 
end of the loop that takes in all the variables. The dummy procedure’s 
pre- and post-conditions will be identical and will represent the 
potential loop invariants. To formally verify Daikon's proposal, 
we inject \texttt{Contracts.Assert} statements for each candidate 
invariant at the beginning and end of the loop. We then invoke the 
static checker provided by the Microsoft Code Contract 
Utility~\cite{fahndrich2010static} to perform the formal verification 
(we made our best efforts to translate programs written in Java and 
Python to C\#). Any loop invariant that can not be proved, albeit may 
still be legitimate, will be removed from the dataset. 
After obtaining loop invariants for positive examples, we 
also generate negative examples to balance the dataset. In particular, 
we exhaustively mutate loop invariants at the token level and pick a 
mutant that is confirmed to be a non-invariant (via random testing) as 
a negative example. Take the program in Figure~\ref{fig:inv} for 
example, after \texttt{max >= diff} is verified to be an loop invariant for the inner loop, 
we mutate the binary operator \texttt{>=} to generate a negative example 
\texttt{max > diff}. Due to the limited power of the static checker, we collected 14,412 
formally verified loop invariants out of 9,663 loops. Along with the same number of 
non loop invariants, we split them into a training set of 20,824, a validation set of 4,000 and a test 
set of the remaining 4,000.

\begin{figure}
	\begin{floatrow}

		\ffigbox{%
			\lstset{style=mystyle}	
\begin{lstlisting}[basicstyle=\linespread{.64}\fontsize{6}{10.8}\ttfamily\bfseries,upquote=true]
static int Difference(int[] a) {
    int max = 0;

    for (int i = 0; i < a.Length-1; i++) { 
        for (int j = i + 1; j < a.Length; j++) { 
            int diff = Math.Abs(a[i] - a[j]);
            max = Math.Max(max, diff);
    }}

    return max;
}
\end{lstlisting}			
		}{%
		    \vspace{-.45cm}
			\caption{Example program for detecting loop invariants.}\label{fig:inv}%
		}
%		\hspace{-.5cm}	
		\capbtabbox{%	
			\resizebox{.45\textwidth}{!}{%
				\bgroup
				\def\arraystretch{.8}
				\begin{tabular}{c | c | c} 
					\hline
					\scriptsize{\textbf{Models}}
					& \scriptsize{\textbf{Accuracy}}
					& \scriptsize{\textbf{F1 Score}}
					\\
					\hline
					\scriptsize{GGNN}  &\scriptsize{47.5\%} &\scriptsize{0.47} \\										
					\hline
					\scriptsize{TreeLSTM} &\scriptsize{51.8\%} &\scriptsize{0.50}\\								
					\hline
					\hline
					\scriptsize{\textbf{\dypro}} &\scriptsize{\textbf{73.9\%}} &\scriptsize{\textbf{0.71}} \\
					\hline
				\end{tabular}
				\egroup
			}
		}{%
			\caption{Classification results for all models in detecting loop invariants.}
			\label{Table:Seres}
		}
	\end{floatrow}
%				\vspace{-.35cm}

\end{figure}

\paragraph{Model Design}
%We depict the dataset in Table~\ref{Table:Sedata}. 
We adapt the existing models to detect loop invariants. To 
establish a fair comparison, we adopt a unified architecture 
across all deep neural networks. Specifically, given a program 
along with the candidate invariants, we concatenate the 
program embedding with the invariants embeddings before 
feeding the concatenations to the prediction layer. At a lower-level, 
we extract embeddings of root nodes in ASTs to be the 
program embeddings for GGNN and TreeLSTM. Regarding loop 
invariants, we feed the token sequences (in the vector format 
according to the embedding matrix of the vocabulary) into another 
RNN from which we extract the final hidden states to be the 
invariant embeddings. Tailoring \dypro towards the unified 
architecture is slightly more involved. First, to embed loop 
invariants as explained above, we incorporate their syntactic 
tokens that are ignored by the state encoding scheme into 
\dypro's vocabulary. Next, for variables appearing in loop invariants, 
we inform \dypro of their values along each step of the execution. In 
particular, we inject the embeddings of the variable Ids to that of their 
values in each program state. We keep the rest of \dypro's architecture 
intact and extract the program embeddings from the pooling layer. We 
experiment two loss functions: cross-entropy loss and hinge loss, and 
find the latter gives better results for all models. As depicted in Table~
\ref{Table:Seres}, \dypro outperforms all static models by a wide margin. 
This strongly indicates \dypro is capable of learning deeper semantic 
properties from executions. On the contrary, by considering the source 
code only, static models fail to learn any plausible mapping from the 
syntactic features to the semantic properties which further confirms static 
models only capture simple, and shallow semantic features.

\subsection{Remarks}
Through two experiments, we have thoroughly demonstrated 
how \dypro stacks up against the static models. To summarize, 
despite the considerable progress, static models are still limited 
in learning semantic representation of a program. It is evident 
their generalization mostly happens at the level of program syntax, 
manifested in their struggle of digging deeper semantic properties. 
In comparison to the static models, \dypro learns from executions, 
a more direct and concrete expression of program semantics. To 
elevate the learning from the level of executions to programs, \dypro 
learns a representation for each individual execution obtained via 
random testing before compressing them into a compact program 
representation. To deal with long execution traces, we equip \dypro 
with the state reduction mechanism so it only generalizes from the 
key program states while discarding the peripherals. Also worth noting, 
even if we did not demonstrate the utility of \dypro in specific problem 
settings, given its strong capability of learning scalable and precise 
representation of program semantics, \dypro should be readily 
applicable to many downstream programming tasks such as bug 
prediction or patch generation.

\section{Conclusion}
\label{sec:con}

In this paper, we propose \dypro, a novel deep 
neural architecture that learns program semantics 
from execution trace. We thoroughly evaluate \dypro 
in a couple of semantically related tasks including 
head-to-head comparisons against several prominent 
static models. Results show \dypro is the most accurate 
in both tasks and more importantly can capture deep semantic properties 
that static models struggle with. For future work, 
we will deploy \dypro to specific problem settings. 
%programming tasks such as bug prediction or program repair. 
Due to its high efficiency and precision in representing the 
program semantics, we expect \dypro to be useful 
in those settings too.

\clearpage
\bibliography{references}

\end{document}